\documentclass[epj]{svjour}
\usepackage{graphics}
\begin{document}

\title{Measurements of the absolute branching fractions for
$D \to \overline K\pi e^+\nu_e$, $D \to \overline K^* e^+\nu_e$ and
determination of $\Gamma(D^+ \to \overline K^{*0}e^+\nu_e)/\Gamma(D^+ \to
\overline K^0 e^+\nu_e)$}
\author{
(BES Collaboration)\\
M.~Ablikim$^{1}$,              J.~Z.~Bai$^{1}$,               Y.~Ban$^{12}$,
J.~G.~Bian$^{1}$,              X.~Cai$^{1}$,
H.~F.~Chen$^{16}$,
H.~S.~Chen$^{1}$,              H.~X.~Chen$^{1}$,              J.~C.~Chen$^{1}$,
Jin~Chen$^{1}$
% ...
}

\institute{
$^{1}$ Institute of High Energy Physics, Beijing 100049, People's Republic of China\\
$^{2}$ China Center for Advanced Science and Technology(CCAST), Beijing 100080, People's Republic of China\\
$^{3}$ Guangxi Normal University, Guilin 541004, People's Republic of China\\
$^{4}$ Guangxi University, Nanning 530004, People's Republic of China\\
$^{5}$ Henan Normal University, Xinxiang 453002, People's Republic of China\\
$^{6}$ Huazhong Normal University, Wuhan 430079, People's Republic of China\\
$^{7}$ Hunan University, Changsha 410082, People's Republic of China\\
$^{8}$ Jinan University, Jinan 250022, People's Republic of China\\
$^{9}$ Liaoning University, Shenyang 110036, People's Republic of China\\
$^{10}$ Nanjing Normal University, Nanjing 210097, People's Republic of China\\
$^{11}$ Nankai University, Tianjin 300071, People's Republic of China\\
$^{12}$ Peking University, Beijing 100871, People's Republic of China\\
$^{13}$ Shandong University, Jinan 250100, People's Republic of China\\
$^{14}$ Sichuan University, Chengdu 610064, People's Republic of China\\
$^{15}$ Tsinghua University, Beijing 100084, People's Republic of China\\
$^{16}$ University of Science and Technology of China, Hefei 230026, People's Republic of China\\
$^{17}$ Wuhan University, Wuhan 430072, People's Republic of China\\
$^{18}$ Zhejiang University, Hangzhou 310028, People's Republic of China
\vspace{0.2cm}\\
$^{a}$ Current address: Purdue University, West Lafayette, IN 47907, USA\\
$^{b}$ Current address: Laboratoire de l'Acc{\'e}l{\'e}rateur Lin{\'e}aire, Orsay, F-91898, France\\
$^{c}$ Current address: University of Michigan, Ann Arbor, MI 48109, USA\\
$^{d}$ Current address: DESY, D-22607, Hamburg, Germany
}

\date{Received: date / Revised version: date}

\abstract{
Using the data of about 33 pb$^{-1}$ collected at and around 3.773 GeV with the BES-II
detector at the BEPC collider, we have studied the exclusive semileptonic decays
$D^+ \to K^-\pi^+ e^+\nu_e$, $D^0 \to \overline K^0\pi^-e^+\nu_e$, $D^+ \to
\overline K^{*0}e^+\nu_e$ and $D^0 \to K^{*-}e^+\nu_e$.
The absolute branching fractions for the decays are measured to be
${BF}(D^+ \to K^-\pi^+e^+\nu_e)=(3.50\pm0.75\pm0.27)\%$,
${BF}(D^0 \to \overline K^0\pi^-e^+\nu_e)=(2.61\pm1.04\pm0.28)\%$,
${BF}(D^+ \to \overline K^{*0}e^+\nu_e)=(5.06\pm1.21\pm0.40)\%$ and
${BF}(D^0 \to K^{*-}e^+\nu_e)=(2.87\pm 1.48\pm 0.39)\%$.
The ratio of the vector to pseudoscalar semileptonic decay rates
$\Gamma(D^+\to\overline K^{*0}e^+\nu_e)/\Gamma(D^+\to\overline K^0e^+\nu_e)$
is determined to be $0.57\pm0.17\pm0.02$.
}

\vspace{-2.0cm}
\maketitle

\section{Introduction}
~~~~Semileptonic decays offer access to weak interaction matrix elements
since the effects of weak and strong interactions can be separated
reasonable well. Measurements of branching fractions for exclusive
decays of $D$ mesons play an important role to develop and to test models
of their decay mechanisms.

Earlier theoretical predictions~\cite{wirbel}\cite{lubicz} implied
that the ratio of the vector to pseudoscalar $D$ meson semileptonic decay rates 
$R=\Gamma(D \to \overline K^* e^+ \nu_e)/\Gamma(D \to
\overline K e^+ \nu_e)$ lies in the range from 0.9 to 1.2~\cite{richman}.
E691 Collaboration reported a measurement with a lower ratio
$\Gamma(D^+ \to \overline K^{*0} e^+ \nu_e)/
   \Gamma(D^0 $\\$\to K^- e^+ \nu_e) = 0.45\pm 0.11$~\cite{e691},
while MARK-III Collaboration obtained
$\Gamma(D \to \overline K\pi e^+ \nu_e)/
   \Gamma(D \to \overline K e^+ \nu_e) = 1.0^{+0.3}_{-0.2}$~\cite{mark3}
which is consistent with the early theoretical predictions.
Since then, some controversies evolved concerning the $R$ ratio from
both theoretical predictions and experimental measurements.
Contrary to the earlier expectations, recent theoretical
predictions and experimental measurements tend to support
a smaller value~\cite{focus}\cite{cleo}. A compilation of
predictions and experimental values can be found in~\cite{focus}.
The $R$ ratio is governed by form factors of the hadronic currents.
So a measurement of $R$ is essential to test form factor calculations.

In this paper, we ~report ~direct ~measurements ~of ~the ~branching ~fractions
~for ~the ~decays ~$D^+ \to K^-\pi^+e^+\nu_e$
(throughout this paper, charge
conjugation is implied), $D^0 \to \overline K^0 \pi^-e^+\nu_e$, $D^+ \to
\overline K^{*0} e^+\nu_e$ and $D^0 \to K^{*-} e^+\nu_e$ by analyzing the
data sample of about 33 pb$^{-1}$~\cite{c5}\cite{c6} collected at and
around the center-of-mass energy 3.773 GeV with the BES-II detector at the
BEPC collider. 

\label{intro}

\section{BES-II detector}
~~~~The BES-II is a conventional cylindrical magnetic detector that is
described in detail in Ref.~\cite {c7}. A 12-layer Vertex Chamber(VC) surrounding
a beryllium beam pipe provides input to event trigger, as well as
coordinate information. A forty-layer main drift chamber (MDC) located
just outside the VC yields precise measurements of charged
particle trajectories with a solid angle coverage of $85\%$ of 4$\pi$;
it also provides ionization energy loss ($dE/dx$) measurements
for particle identification. Momentum
resolution of $1.7\%\sqrt{1+p^2}$ ($p$ in GeV/$c$) and $dE/dx$
resolution of $8.5\%$ for Bhabha scattering electrons are obtained for
the data taken at $\sqrt{s}=3.773$ GeV. An array of 48 scintillation
counters surrounding the MDC measures time of flight (TOF) of
charged particles with a resolution of about 180 ps for electrons.
Outside the TOF counters, a 12 radiation length, lead-gas barrel shower counter
(BSC), operating in limited streamer mode, measures the energies of
electrons and photons over $80\%$ of the total solid angle with an
energy resolution of $\sigma_E/E=0.22/\sqrt{E}$ ($E$ in GeV) and spatial
resolutions of $\sigma_{\phi}=7.9$ mrad and $\sigma_Z=2.3$ cm for
electrons. A solenoidal magnet outside the BSC provides a 0.4 T
magnetic field in the central tracking region of the detector. Three
double-layer muon counters instrument the magnet flux return and serve
to identify muons with momentum greater than 500 MeV/c. They cover
$68\%$ of the total solid angle.

\section{\normalsize \bf Data analysis}
The $\psi(3770)$ resonance is produced in electron-positron ($e^+e^-$)
annihilation at the center-of-mass energy of about 3.773 GeV.
It is believed to decay predominately into $D^0\bar D^0$ and $D^+D^-$
pairs. Therefore, if a $\bar D$ meson is fully reconstructed
(this is called a singly tagged $\bar D$ meson)~\cite{c5}\cite{c6},
the $D$ meson must exist in the system recoiling against the singly
tagged $\bar D$ meson. In the system recoiling against singly tagged $D^-$
and $\bar D^0$ mesons, we select semileptonic decays
$D^+ \to K^- \pi^+(\overline K^{*0})e^+\nu_e$ and $D^0\to \overline K^0
\pi^-(K^{*-})e^+\nu_e$ respectively, and measure branching
fractions for the decays directly.

\subsection{\normalsize \bf Event selection}
~~~~In order to ensure the well-measured 3-momentum vectors and the
reliability of the charged-particle identification, all charged tracks
are required to be well reconstructed in the MDC with good helix fits, and to
satisfy a geometry cut $|\rm{cos\theta}|<0.85$, where $\theta$ is the polar
angle. Each track, except for those from $K^0_S$ decays, must originate from the
interaction region, which is defined by $V_{xy}<2.0$ cm and $|V_z|<20.0$ cm,
where $V_{xy}$ and $|V_z|$ are the closest approach of the charged track 
in the $xy$-plane and $z$ direction.

Pions and kaons are identified using the $dE/dx$ and TOF measurements, with
which the combined confidence levels ($CL_{\pi}$ or $CL_K$) for a pion or kaon
hypotheses are calculated. A pion candidate is required to have $CL_{\pi}>
0.001$. In order to reduce misidentification, a kaon candidate is required to
satisfy $CL_K>CL_{\pi}$. Electrons are identified using the $dE/dx$, TOF and
BSC measurements, with which the combined confidence level ($CL_e$) for
the electron hypotheses is calculated. An electron candidate is required to
have $CL_e>0.001$ and satisfy the relation $CL_e/(CL_e+CL_K+CL_{\pi})>0.8$.

Neutral kaons are ~reconstructed ~through ~the ~decay $K^0_S \to \pi^+\pi^-$.
The difference between the invariant mass of $\pi^+\pi^-$ combinations
and the $K^0_S$ nominal mass should be less than $20$ MeV/$c^2$.

Neutral pions are reconstructed through ~the ~decay $\pi^0 \to \gamma\gamma$.
A good photon candidate must satisfy the following criteria: (1) the energy
deposited in the BSC is greater than 70 MeV;
(2) the electromagnetic shower starts in the first 5 readout layers; (3) the angle
between the photon and the nearest charged track is greater than $22^\circ$;
(4) the opening angle between the direction of the cluster development and the
direction of the photon emission is less than $37^\circ$.

\begin{figure}
\resizebox{0.44\textwidth}{0.345\textheight}{%
  \includegraphics{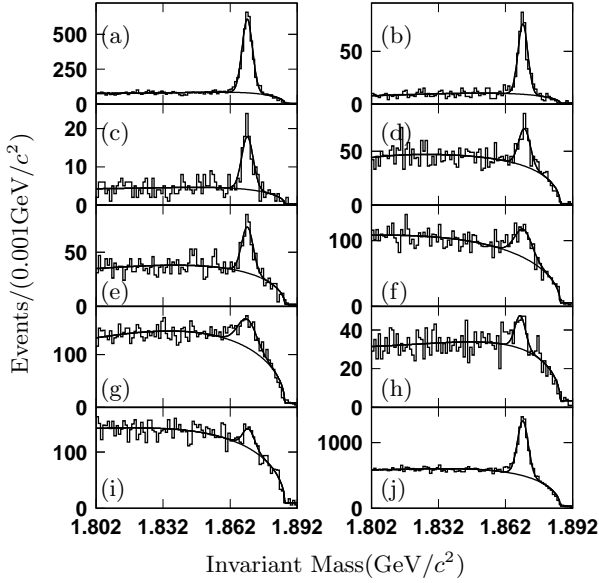}}
\put(-150,5.0){Invariant Mass(GeV/$c^2$)}
\put(-225,80){\rotatebox{90}{Events/(0.001GeV/$c^2$)}}
\put(-190,205){(a)}
\put(-84,205){(b)}
\put(-190,168.5){(c)}
\put(-84,168.5){(d)}
\put(-190,108){(e)}
\put(-84,108){(f)}
\put(-190,71.5){(g)}
\put(-84,71.5){(h)}
\put(-190,33.5){(i)}
\put(-84,33.5){(j)}
\caption{The distributions of the fitted invariant masses of the $nKm\pi
\hspace{0.1cm}
(n=0,1,2; m=1,2,3,4)$ combinations in the singly tagged $D^-$ modes:
(a) $K^+\pi^-\pi^-$, (b) $K^0 \pi^-$,
(c) $K^0 K^-$,
(d) $K^+ K^-\pi^-$,
(e) $K^0\pi^-\pi^-\pi^+$,
(f) $K^0\pi^-\pi^0$,
(g) $K^+\pi^-\pi^-\pi^0$,
(h) $K^+\pi^-\pi^-\pi^-\pi^+$,
(i) $\pi^-\pi^-\pi^+$ and
(j) sum of the nine modes.}
\label{dp9stag}
\end{figure}

\begin{figure}
\resizebox{0.44\textwidth}{0.345\textheight}{%
  \includegraphics{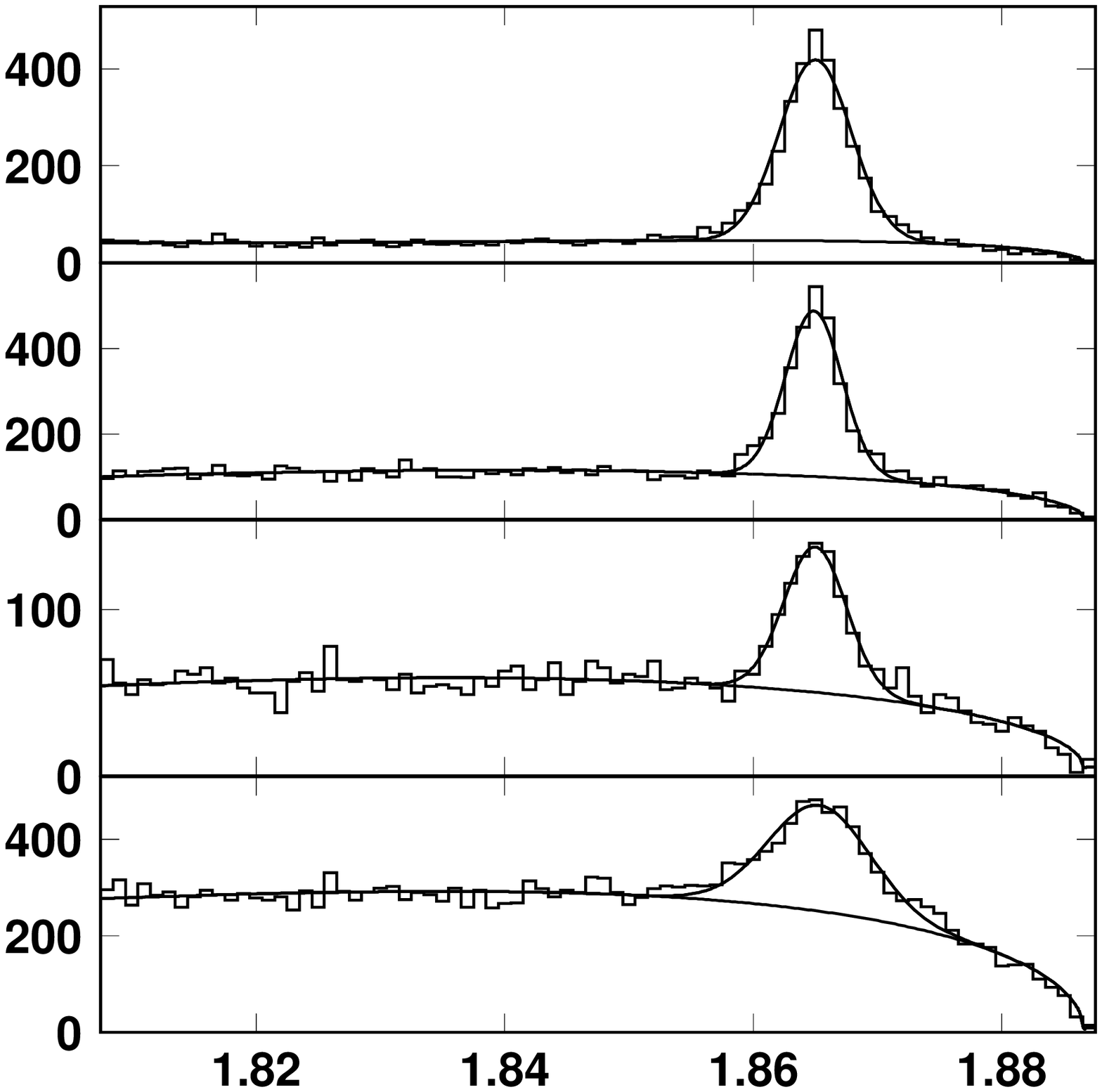}}
\put(-150,5.0){Invariant Mass(GeV/$c^2$)}
\put(-225,80){\rotatebox{90}{Events/(0.001GeV/$c^2$)}}
\put(-180,205){(a)}
\put(-180,157.5){(b)}
\put(-180,110.0){(c)}
\put(-180,62.5){(d)}
\caption{
The distributions of the fitted invariant masses of the $Km\pi
\hspace{0.1cm}
(m=1,2,3)$ combinations in the singly tagged $\bar D^0$ modes:
(a) $ K^+\pi^-$, (b) $ K^+\pi^-\pi^-\pi^+$,
(c) $ K^0\pi^+\pi^-$ and (d) $K^+\pi^-\pi ^0$.}
\label{d04stag}
\end{figure}

\subsection{\normalsize \bf Singly tagged $D^-$ and $\bar D^0$ samples}
~~~~The singly tagged $D^-$ and $\bar D^0$ samples used in this analysis were
selected in the previous work~\cite{c5}\cite{c6}, we here give a brief description
for the selection of the singly tagged $D^-$ and $\bar D^0$ samples.

The singly tagged $D^-$ mesons are reconstructed in nine hadronic decay
modes of $K^+\pi^-\pi^-$, $K^0 \pi^-$, $K^0 K^-$, ~~ $K^+ K^-\pi^-$, ~~ $K^0
\pi^-\pi^-\pi^+$, ~~ $K^0\pi^-\pi^0$, ~~ $K^+\pi^-\pi^-\pi^0$, ~$K^+\pi^-\pi^-
\pi^-\pi^+$ and $\pi^-\pi^-\pi^+$. And the singly tagged $\bar D^0$ mesons
are reconstructed in four hadronic decay modes of $K^+\pi^-$, $K^+\pi^-
\pi^-\pi^+$, $K^0\pi^+\pi^-$ and $K^+\pi^-\pi^0$.

In order to improve the momentum resolution and the ratio of signal to
combinatorial background in the invariant mass spectra, the center-of-mass energy constraint
kinematic fit is imposed on each of the $nKm\pi \hspace{0.1cm}
(n=0,1,2;$\\$m=1,2,3,4)$
combinations. If there is a $K_S^0$ or $\pi^0$ among the $D$ daughter particles,
an additional constraint kinematic fit will be imposed on the decay 
$K_S^0 \to \pi^+\pi^-$ or $\pi^0 \to \gamma \gamma$. Combinations with a
kinematic fit probability greater than $0.1\%$ are accepted. If more than
one combination satisfies the criteria in an event, only the combination with
the largest fit probability is retained.

The resulting distributions of the fitted invariant masses of the
$nKm\pi$ combinations, which are calculated
using the fitted momentum vectors from the kinematic fit, are shown in Fig.~1 and
Fig.~2 for the singly tagged $D^-$ and $\bar D^0$ modes, respectively.
A maximum likelihood fit to the mass spectrum with a Gaussian function for
the $\bar D$ signal and a special function~\cite{c5}\cite{c6} to describe
the background shape yields the observed numbers of the singly tagged $D^-$ and
$\bar D^0$ mesons for each mode. These give the total number of the
reconstructed singly tagged $\bar D$ mesons, $5321\pm149\pm160$ for
$D^-$~\cite{c5} and $7584\pm198\pm341$ for $\bar D^0$~\cite{c6}, where the first
error is statistical and the second systematic obtained by varying
the parameterization of the background.

\subsection{Candidates for $D^+ \to K^-\pi^+e^+\nu_e$
and $D^0 \to \overline K^0 \pi^-e^+\nu_e$}

~~~~Candidates for the semileptonic decays $D^+ \to K^-\pi^+e^+\nu_e$ and
$D^0 \to \overline K^0 \pi^-e^+\nu_e$ are selected from the surviving
tracks in the system recoiling against the singly tagged $\bar D$ mesons.
For the selected candidate events,
it is required that there should be no extra charged track
or isolated photon, which has not been used in the reconstruction of the singly
tagged $\bar D$ mesons. The isolated photon detected in the BSC should have
an energy exceeding 100 MeV and should satisfy the photon selection criteria
described earlier. There are possible hadronic backgrounds for each semileptonic
decay due to misidentification of a charged pion as an electron, for example,
the decay 
$D^+ \to K^-\pi^+\pi^+$ ($D^0 \to \overline K^0 \pi^+\pi^-$) 
could be misidentified as $D^+ \to K^- \pi^+e^+\nu_e$
($D^0 \to \overline K^0 \pi^-e^+\nu_e$). However, 
these events can be suppressed by requiring the
invariant masses of $K^-\pi^+e^+$ ($\overline K^0\pi^-e^+$) combinations to be less than
1.75 GeV/$c^2$.

\begin{figure}
\resizebox{0.44\textwidth}{0.345\textheight}{%
  \includegraphics{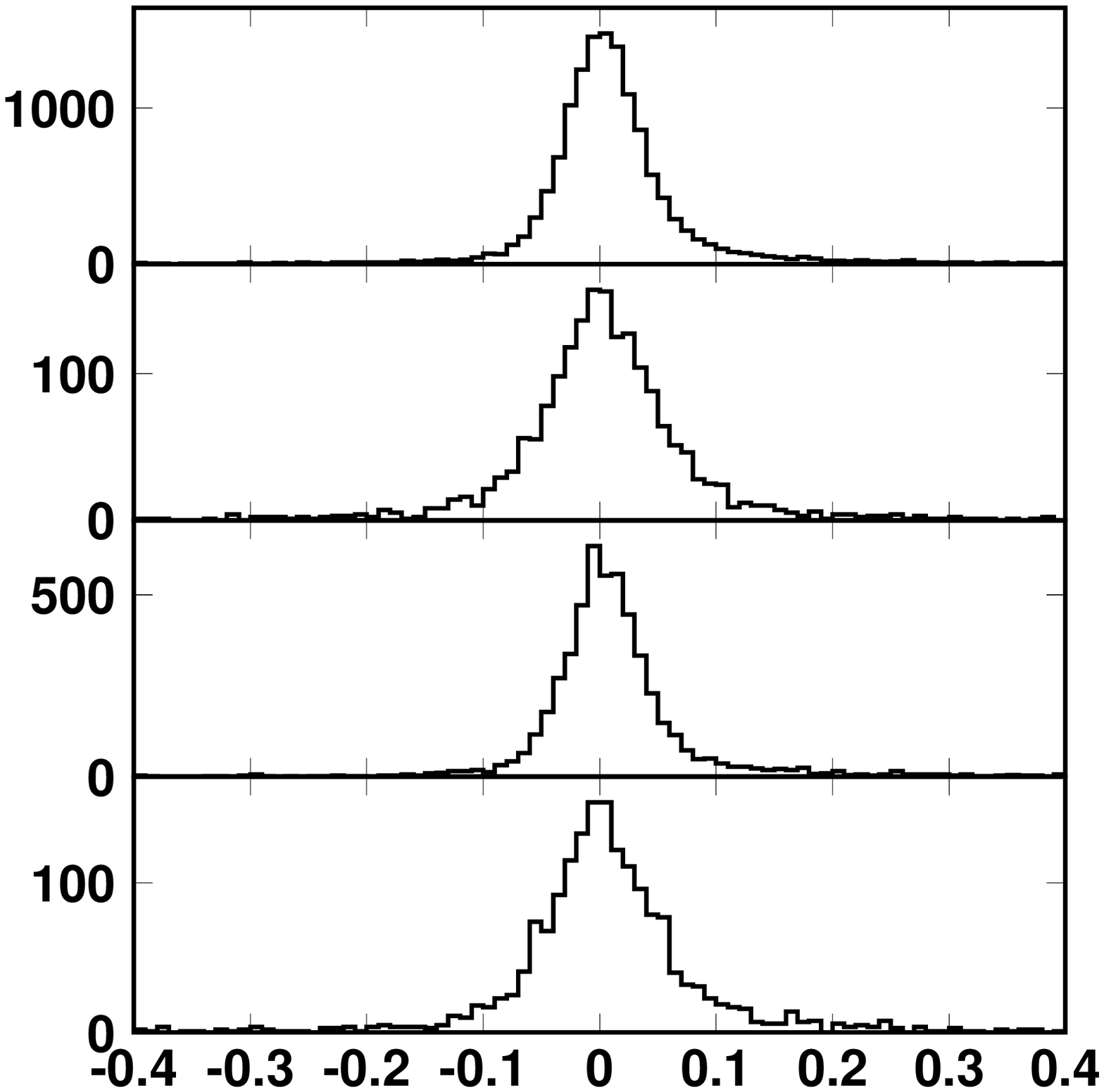}}
\put(-130,5.0){$U_{miss}$(GeV)}
\put(-225,85){\rotatebox{90}{Events/(0.01GeV)}}
\put(-180,205){(a)}
\put(-180,157.5){(b)}
\put(-180,110.0){(c)}
\put(-180,62.5){(d)}
\caption{
The distributions of the $U_{miss}$ for the Monte Carlo events of
(a) $D^+ \to K^-\pi^+e^+\nu_e$ versus $D^- \to K^+\pi^-\pi^-$,
(b) $D^+ \to K^-\pi^+e^+\nu_e$ versus $D^- \to K^+\pi^-\pi^-\pi^0$,
(c) $D^0 \to \overline K^0 \pi^-e^+\nu_e$ versus $\bar D^0 \to K^+\pi^-$ and
(d) $D^0 \to \overline K^0 \pi^-e^+\nu_e$ versus $\bar D^0 \to
K^+\pi^-\pi^0$.}
\label{umissmc}
\end{figure}

\begin{figure}
\resizebox{0.44\textwidth}{0.345\textheight}{%
  \includegraphics{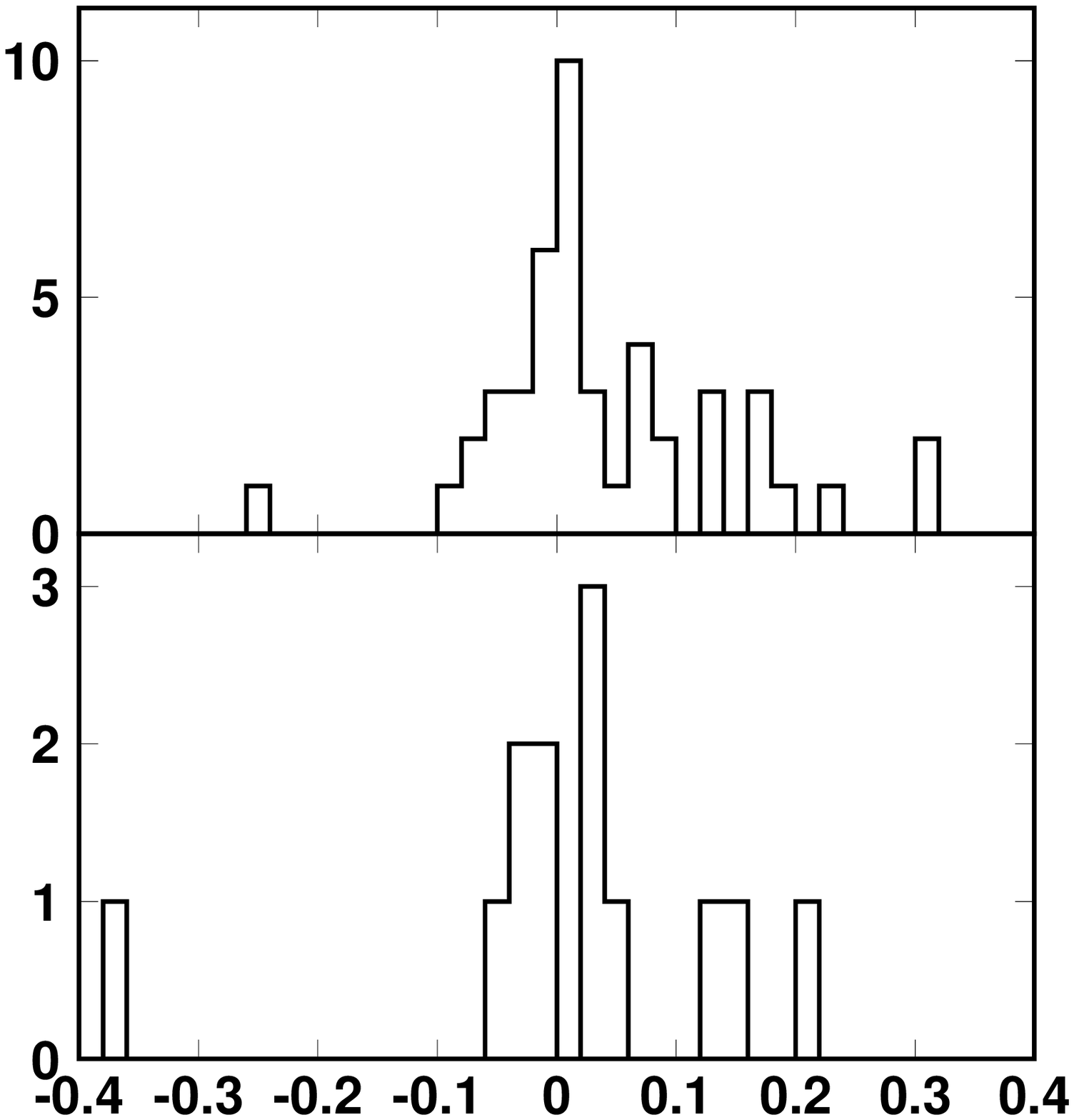}}
\put(-130,5.0){$U_{miss}$(GeV)}
\put(-225,85){\rotatebox{90}{Events/(0.02GeV)}}
\put(-180,205){(a)}
\put(-180,110.0){(b)}
\caption{
The distributions of the $U_{miss}$ for the selected candidates for
(a) $D^+ \to K^-\pi^+e^+\nu_e$ and
(b) $D^0 \to \overline K^0 \pi^-e^+\nu_e$ from the data.}
\label{umissdata}
\end{figure}

In the semileptonic decays, there is one massless neutrino which is undetected.
To obtain information about missing neutrinos, a kinematic quantity $U_{miss}
\equiv E_{miss} - p_{miss}$ is used~\cite{c5}\cite{c6}, where $E_{miss}$
and $p_{miss}$ are the total energy and momentum of all missing particles.
To select the semileptonic decays,
it is required that the candidates for the semileptonic decays should have
their $|U_{miss,i}| <3\sigma_{U_{miss,i}}$, where the $\sigma_{U_{miss,i}}$
is the standard deviation of the $U_{miss,i}$ distribution 
obtained by analyzing the Monte Carlo events
of $D^+ \to K^-\pi^+e^+\nu_e$ ($D^0 \to \overline K^0\pi^- e^+ \nu_e$) versus
the $i$th singly tagged $\bar D$ mode. Fig.~3 shows the distributions of the
$U_{miss}$, with each peak centered at zero as expected, for the Monte Carlo events of
$D^+ \to K^-\pi^+e^+\nu_e$ versus $D^- \to K^+\pi^-\pi^-$,
$D^+ \to K^-\pi^+e^+\nu_e$ versus $D^- \to K^+\pi^-\pi^-\pi^0$,
$D^0 \to \overline K^0 \pi^-e^+\nu_e$ versus $\bar D^0 \to K^+\pi^-$
and $D^0 \to \overline K^0 \pi^-e^+\nu_e$ versus $\bar D^0 \to
K^+\pi^-\pi^0$, respectively. Fig.~4(a) and Fig.~4(b) respectively show
the distributions of the $U_{miss}$ for the selected candidates for
$D^+ \to K^-\pi^+e^+\nu_e$ and $D^0 \to \overline K^0 \pi^-e^+\nu_e$
from the data.

\begin{figure}
\resizebox{0.44\textwidth}{0.345\textheight}{%
  \includegraphics{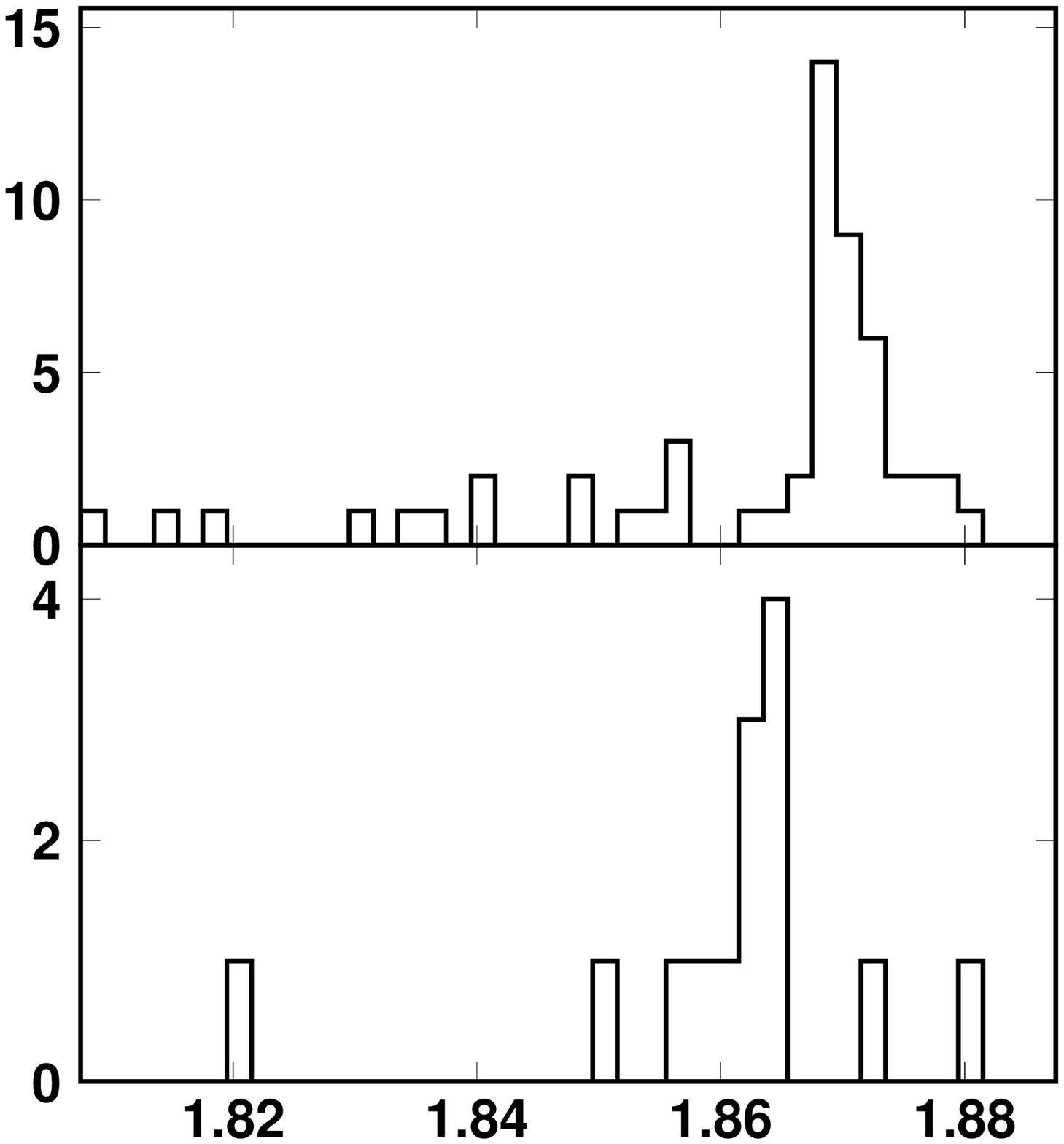}}
\put(-160,5.0){Invariant Mass(GeV/$c^2$)}
\put(-225,85){\rotatebox{90}{Events/(0.002GeV/$c^2$)}}
\put(-180,205){(a)}
\put(-180,110.0){(b)}
\caption{
The distributions of the fitted invariant masses of the $nKm\pi$
combinations
for the events in which the candidates for
(a) $D^+ \to K^-\pi^+e^+\nu_e$ and (b) $D^0 \to \overline K^0 \pi^-e^+\nu_e$
are observed in the system recoiling against the singly tagged $\bar D$.
}
\label{dkpiev}
\end{figure}

Figure~5 shows the distributions of the fitted invariant masses of the $nKm\pi$
combinations from the events in which the candidates for $D^+ \to K^-\pi^+e^+
\nu_e$ and $D^0 \to \overline K^0\pi^-e^+\nu_e$ are selected in the system
recoiling against the $nKm\pi$ combinations. Fig.~5(a) indicates that an
obvious signal of $D^+ \to K^-\pi^+e^+\nu_e$ via the $D^-$ tags is observed.
There are 37 events in the $\pm 3\sigma_{M_{D_i}}$ mass window of the fitted
$D$ meson mass $M_{D_i}$, and 18 events in the outside of the signal regions,
where $\sigma_{M_{D_i}}$ is the standard deviation of the $nKm\pi$ distribution
for the $i$th mode. By assuming that the distribution of background is flat,
$5.6\pm1.4$ background events in the signal regions are estimated.
After subtracting the number of background events,
$31.4\pm6.2$ candidates for $D^+ \to K^-\pi^+e^+\nu_e$ are retained. A similar
analysis of the events in Fig.~5(b) yields $9.9\pm3.4$ candidates for 
$D^0 \to \overline K^0 \pi^-e^+\nu_e$. There may also be the $\pi^+\pi^-$
combinatorial background. By selecting the events in which the invariant
masses of the $\pi^+\pi^-$ combinations on the recoil side of the tags
are outside of the $K^0_S$ mass window, we estimate that there are
$0.6\pm0.2$ background events in the candidate events. After subtracting
the number of background events, $9.3\pm3.4$ candidate events are retained.

\subsection{\bf \normalsize Candidates for $D^+ \to
\overline K^{*0}e^+\nu_e$ and $D^0 \to K^{*-}e^+\nu_e$}

~~~~To select the candidates for $D^+ \to \overline K^{*0}e^+\nu_e$ and
$D^0 \to K^{*-}e^+\nu_e$, we calculate the invariant masses of
$K^-\pi^+$ ($\overline K^0\pi^-$) combinations from the selected candidates for
$D^+\to K^-\pi^+e^+\nu_e$ ($D^0\to \overline K^0\pi^-e^+\nu_e$). Fig.~6(a)
and Fig.~6(b) show the distributions of the invariant masses of
$K^-\pi^+$ and $\overline K^0\pi^-$ combinations for the events in which
the invariant masses of the $nKm\pi$ combinations are within the
$\pm 3\sigma_{M_{D_i}}$ mass window of the fitted $D$ meson mass
$M_{D_i}$. A clear $\overline K^{*0}$
signal is observed in Fig.~6(a). Fitting the $K^-\pi^+$ invariant mass
spectrum with a Gaussian function for the $\overline K^{*0}$ signal and a
S-wave $K\pi$ phase space background shape,
we obtain $29.1\pm6.6$ candidates
for $D^+ \to\overline K^{*0} e^+\nu_e$. In the fit, the mass and width of
$\overline K^{*0}$ are respectively fixed to 0.8961 GeV/$c^2$ and 50.7 MeV/$c^2$ quoted
from PDG~\cite{c8}, the mass resolution is fixed to 10 MeV/$c^2$ determined
by Monte Carlo simulation.
Using the same background shape, a similar fit to the $\overline K^0\pi^-$
invariant mass spectrum in Fig.~6(b) yields
$7.4\pm3.3$ candidates for $D^0 \to K^{*-}e^+\nu_e$. After subtracting the
number of $\pi^+\pi^-$ combinatorial background of $0.1\pm0.1$ events, $7.3\pm3.3$
candidate events are retained.

\begin{figure}
\resizebox{0.44\textwidth}{0.345\textheight}{%
  \includegraphics{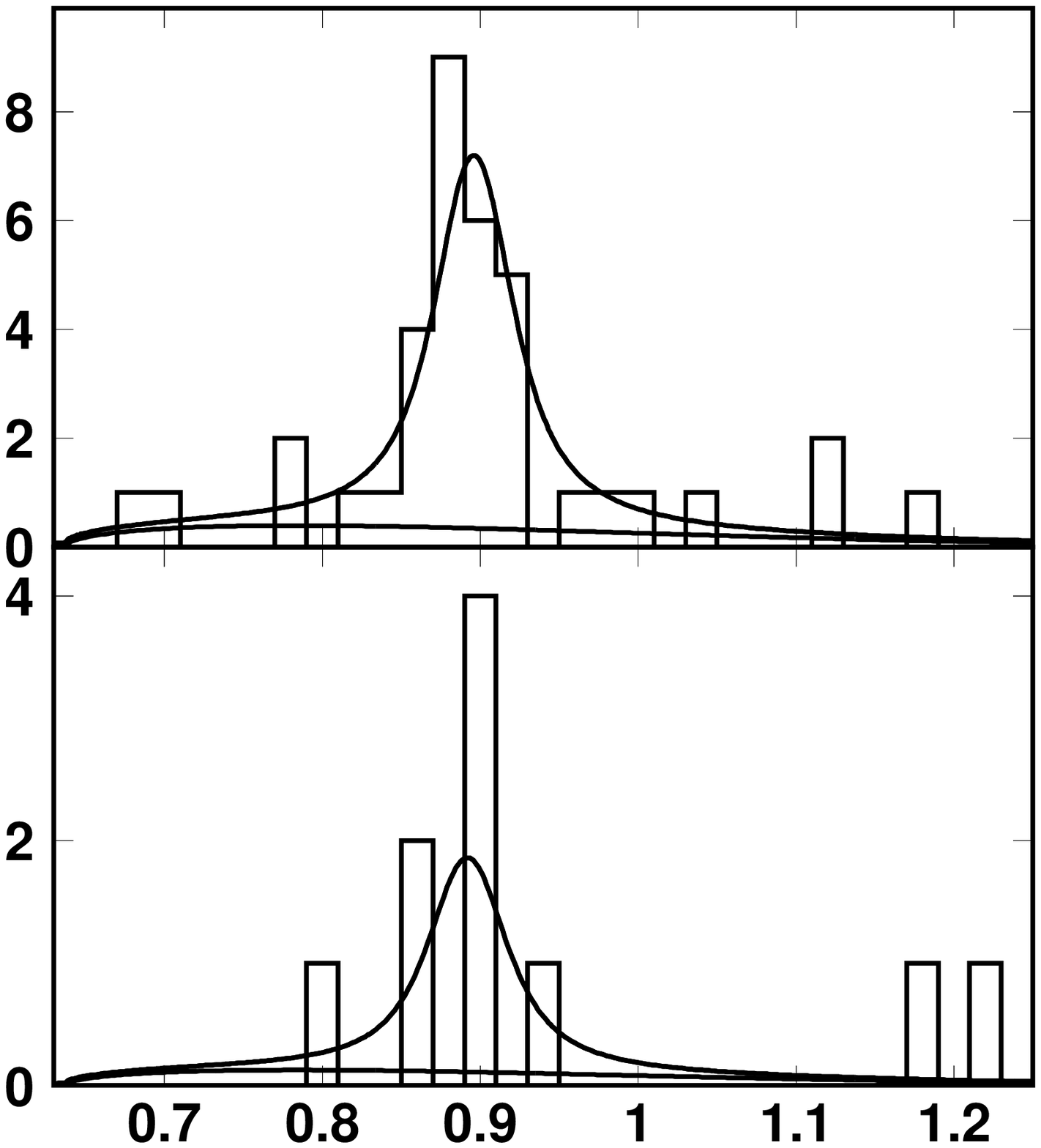}}
\put(-160,5.0){Invariant Mass(GeV/$c^2$)}
\put(-225,80){\rotatebox{90}{Events/(0.02GeV/$c^2$)}}
\put(-180,205){(a)}
\put(-180,110.0){(b)}
\caption{
The distributions of the invariant masses of (a) $K^-\pi^+$ combinations
from the selected candidates for $D^+ \to \overline K^{*0}e^+\nu_e$ and (b)
$\overline K^0\pi^-$ combinations
from the selected candidates for $D^0 \to K^{*-}e^+\nu_e$.
}
\label{dkstarev}
\end{figure}

However, there are still some $K^*$ contaminations from other modes
of $D$ meson decays or from continuum background due to the combinatorial
background in the singly tagged $\bar D$ signal regions. These $K^*$
contaminations must be subtracted from the fitted number of the selected
candidates for $D^+ \to \overline K^{*0}e^+\nu_e$ and $D^0 \to K^{*-}e^+\nu_e$.
They are estimated by using the $\bar D$ sideband events.
The numbers of the events satisfying the
selection criteria in the $\bar D$ sideband are then normalized to
obtain the numbers of the background events in the $\bar D$ signal regions.
Totally $0.8\pm0.8$ and $0.7\pm0.5$ background events for $D^+\to\overline K^{*0}e^+\nu_e$
and $D^0\to K^{*-}e^+\nu_e$ are obtained respectively. After subtracting the numbers of
the background events, $28.3\pm6.6$ and $6.6\pm3.3$ candidate events for
$D^+ \to \overline K^{*0}e^+\nu_e$ and $D^0 \to K^{*-}e^+\nu_e$ are
respectively retained.

The distribution of the momentum of the electrons from the selected candidates for
$D^+ \to \overline K^{*0}e^+\nu_e$ is shown in Fig.~7, where the 
points with error bars are from the data 
and the histogram is from the
Monte Carlo events of $D^+ \to \overline K^{*0}e^+\nu_e$.
The contaminations from hadronic and some other semileptonic decays have been
subtracted from the observed events based on study of the Monte Carlo events
of $e^+e^-\to D \bar D$. 

\begin{figure}
\resizebox{0.44\textwidth}{0.345\textheight}{%
  \includegraphics{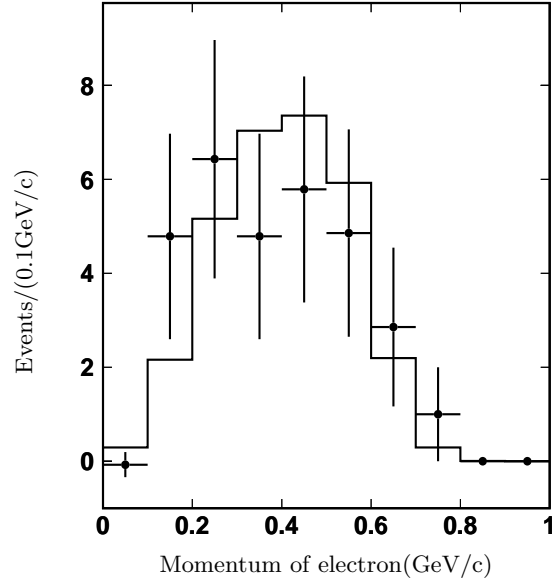}}
\put(-170,5.0){Momentum of electron(GeV/c)}
\put(-225,80){\rotatebox{90}{Events/(0.1GeV/c)}}
\caption{
The distribution of the momentum of the electrons from the selected
candidates for
$D^+ \to \overline K^{*0}e^+\nu_e$, where the points with error bars are
from the data and histogram is from the Monte
Carlo events of $D^+ \to \overline K^{*0}e^+\nu_e$.
}
\label{pesemi}
\end{figure}

\subsection{\bf \normalsize Other backgrounds}
~~~~The events from other hadronic or semileptonic decays may also
satisfy the selection criteria for the semileptonic decays and
are misidentified as the semileptonic decay events. The numbers of
the misidentified events have to be subtracted from the candidates for the
semileptonic decays. The numbers of the background events
are estimated by analyzing the Monte Carlo sample which is about
14 times larger than the data. The Monte Carlo events are generated as
$e^+e^- \to D\bar D$, where the $D$ and $\bar D$ mesons are set to decay
into all possible final states with the branching fractions quoted from
PDG~\cite{c8} excluding the decay modes under study. The particle trajectories are
simulated with the GEANT3 based Monte Carlo simulation package for the
BES-II detector~\cite{c9}. The number of the events satisfying the selection
criteria is then normalized to the data. Monte Carlo study shows that
the dominant background for $D^+ \to K^-\pi^+(\overline K^{*0})e^+\nu_e$
is from $D^+ \to \overline K^{*0}\mu^+\nu_{\mu}$, and the background for
$D^0 \to \overline K^0 \pi^-(K^{*-})e^+\nu_e$ is from $D^0 \to K^{*-}\pi^+$.
Totally $2.5\pm0.5$, $0.8\pm0.3$,
$0.7\pm0.3$ and $0.2\pm0.3$ background events for $D^+ \to K^-\pi^+e^+\nu_e$,
$D^0 \to \overline K^0\pi^-e^+\nu_e$, $D^+ \to\overline K^{*0}e^+\nu_e$ and $D^0 \to
K^{*-}e^+\nu_e$ are obtained, respectively. After subtracting the numbers of
the background events, $28.9\pm6.2$, $8.5\pm3.4$, $27.6\pm6.6$ and $6.4\pm3.3$ signal
events for the semileptonic decays $D^+ \to K^- \pi^+e^+\nu_e$, $D^0 \to
\overline K^0 \pi^-e^+\nu_e$, $D^+ \to \overline K^{*0}e^+\nu_e$ and $D^0\to K^{*-}e^+\nu_e$
are obtained.

\section{\bf \normalsize Results}
\subsection{\bf \normalsize Monte Carlo efficiency}
~~~~The detection efficiencies for the semileptonic decays
$D^+ \to K^-\pi^+e^+\nu_e$, $D^0 \to\overline K^0\pi^-e^+\nu_e$,
$D^+ \to \overline K^{*0}e^+\nu_e$ and $D^0 \to K^{*-}e^+\nu_e$ are
estimated to be 
$\epsilon_{D^+\to K^-\pi^+e^+\nu_e}=(15.51\pm0.12)\%$,
$\epsilon_{D^0 \to \overline K^0\pi^-e^+\nu_e}=(4.30\pm0.05)\%$,
$\epsilon_{D^+ \to \overline K^{*0}e^+\nu_e}$\\$=(10.26\pm0.08)\%$
and $\epsilon_{D^0 \to K^{*-}e^+\nu_e}=(2.94\pm0.04)\%$
by Monte Carlo simulation,
which include the branching fractions for the intermediate sub-resonance decays. 

\subsection{\bf \normalsize Branching fractions}
~~~~The branching fraction for the semileptonic decay $D \to j$ (where $j$ =
$K^-\pi^+e^+\nu_e$, $\overline K^0\pi^-e^+\nu_e$, $\overline K^{*0}e^+\nu_e$
and $K^{*-}e^+\nu_e$) can be determined by
\begin{equation}
{BF}(D \to j)=\frac{N_{D \to j}}
{N_{\bar D_{tag}}\times\epsilon_{D \to j}},
\label{brsemi}
\end{equation}
where $N_{D \to j}$ is the number of the signal events for the $j$th mode;
$N_{\bar D_{tag}}$ is the total number of the singly tagged $D^-$ or $\bar D^0$ mesons;
$\epsilon_{D\to j}$ is the detection efficiency for the $j$th mode.
Inserting these numbers in Eq.~(\ref{brsemi}), we obtain the branching fractions
for the semileptonic decays to be
$${BF}(D^+ \to K^-\pi^+e^+\nu_e)=(3.50\pm0.75\pm0.27)\%,$$
$${BF}(D^0 \to \overline K^0\pi^-e^+\nu_e)=(2.61\pm1.04\pm0.28)\%,$$
$${BF}(D^+ \to \overline K^{*0}e^+\nu_e)=(5.06\pm1.21\pm0.40)\%$$
and $${BF}(D^0 \to K^{*-}e^+\nu_e)=(2.87\pm 1.48\pm 0.39)\%,$$
where the first error is statistical and the second systematic.
The systematic error arises mainly from the uncertainties 
in tracking efficiency ($\sim$2.0\% per track),
in particle identification ($\sim$0.5\% per track 
for charged pion or kaon, $\sim$1.0\% per track for electron), 
in photon selection ($\sim$2.0\%), 
in $K^0_S$ selection ($\sim$1.1\%),
in $U_{miss}$ selection ($\sim$0.6\%), 
in background subtraction [$\sim$(2.5\%$\sim$9.3\%)], 
in Monte Carlo statistics [$\sim$(0.8\%$\sim$1.4\%)],
in the number of the singly tagged $\bar D$ mesons ($\sim$3.0\% for $D^-$ 
and $\sim$4.5\% for $\bar D^0$) and in the fit to the mass spectrum of $K^-\pi^+$ or
$\overline K^0 \pi^-$ combination ($\sim$1.7\% for $D^+\to \overline K^{*0}e^+\nu_e$
and $\sim$2.5\% for $D^0 \to K^{*-}e^+\nu_e$).
These uncertainties are added in quadrature to obtain the total
systematic error, yielding $\sim$7.6\%, $\sim$10.8\%, $\sim$8.0\% and $\sim$13.7\%
for the semileptonic decays $D^+ \to K^-\pi^+e^+\nu_e$, $D^0 \to \overline K^0\pi^-e^+\nu_e$,
$D^+ \to \overline K^{*0}e^+\nu_e$ and $D^0 \to K^{*-}e^+\nu_e$, respectively.

\subsection{\bf \normalsize The ratio of $\frac{\Gamma(D^+ \to \overline K
^{*0}e^+\nu_e)}{\Gamma(D^+ \to \overline K^0 e^+\nu_e)}$}

~~~~With the measured branching fraction for $D^+ \to \overline K^{*0}e^+\nu_e$
and the previously measured branching fraction 
${BF}(D^+ \to \overline K^0 e^+\nu_e)= (8.95\pm1.59\pm0.67)\%$ by BES
Collaboration~\cite{c5}, we obtain the ratio of the vector to pseudoscalar
$D$ meson semileptonic decay rates to be 
$$\frac {\Gamma(D^+ \to\overline K^{*0}e^+\nu_e)}
{\Gamma(D^+ \to \overline K^0 e^+\nu_e)}=0.57\pm0.17\pm0.02,$$
where the first error is statistical and the second systematic which arises
mainly from the uncanceled systematic uncertainties including $K^0_S$ selection
($\sim$1.1\%), background subtraction ($\sim$3.5\%), Monte Carlo statistics
($\sim$1.1\%) and the fit to the mass spectrum of $K^-\pi^+$ combinations
($\sim$1.7\%).

\section{\bf \normalsize Summary}
~~~~Using the data of about 33 pb$^{-1}$ collected around 3.773 GeV with
the BES-II detector at the BEPC collider, the absolute branching fractions for
the decays $D^+ \to K^-\pi^+e^+\nu_e$, $D^0\to \overline K^0\pi^-e^+\nu_e$,
$D^+ \to \overline K^{*0}e^+\nu_e$ and $D^0 \to K^{*-}e^+ \nu_e$ are
measured to be ${BF}(D^+ \to K^-\pi^+e^+\nu_e)=(3.50\pm0.75\pm0.27)\%$,
${BF}(D^0 \to \overline K^0\pi^-e^+\nu_e)=(2.61\pm1.04\pm0.28)\%$,
${BF}(D^+ \to \overline K^{*0}e^+\nu_e)=(5.06\pm1.21\pm0.40)\%$ and
${BF}(D^0 \to K^{*-}e^+\nu_e)=(2.87\pm 1.48\pm 0.39)\%$. With the measured
branching fraction for $D^+ \to \overline K^{*0}e^+\nu_e$ and the previously
measured branching fraction for $D^+ \to \overline K^0 e^+\nu_e$, the ratio
of the vector to pseudoscalar $D$ meson semileptonic decay rates
$\Gamma(D^+\to\overline K^{*0}e^+\nu_e)/\Gamma(D^+\to\overline K^0e^+\nu_e)$
is determined to be
$0.57\pm0.17\pm0.02$, which is in good agreement with theoretical
predictions and other measurements~\cite{focus}\cite{cleo} within error.

\section{Acknowledgment}
~~~~The BES collaboration thanks the staff of BEPC and computing center
for their hard efforts. This work is supported in part by the National
Natural Science Foundation of China under contracts Nos. 10491300,
10225524, 10225525, 10425523, the Chinese Academy of Sciences under
contract No. KJ 95T-03, the 100 Talents Program of CAS under
Contract Nos. U-11, U-24, U-25, and the Knowledge Innovation
Project of CAS under Contract Nos. U-602, U-34 (IHEP), the
National Natural Science Foundation of China under Contract No.
10225522 (Tsinghua University).


\begin{thebibliography}{}
\bibitem{wirbel} M. Wirbel, B. Stech and M. Baucer, Z. Phys. {\bf C} 29 (1985) 637.
\bibitem{lubicz} V. Lubicz et. al, Phys. Lett. {\bf B} 274 (1992) 415.
\bibitem{richman} J.D. Richman and P.R. Burchat, Rev. Mod. Phys. 67 (1995) 893.
\bibitem{e691} The Tagged Photon Spectrometer Collaboration, J.C. Anjos et al.,
Phys. Rev. Lett. 62 (1989) 722.
\bibitem{mark3} Mark III Collaboration, Z. Bai et al.,
Phys. Rev. Lett. 66 (1991) 1011.
\bibitem{focus} FOCUS Collaboration, J.M. Link et al.,
Phys. Lett. {\bf B} 598 (2004) 33.
\bibitem{cleo} CLEO Collaboration, G.S. Huang et al.,
Phys. Rev. Lett. 95 (2005) 181801.
\bibitem{c5} BES Collaboration, M. Ablikim, et al.,
Phys. Lett. {\bf B} 608 (2005) 24.
\bibitem{c6} BES Collaboration, M. Ablikim, et al.,
Phys. Lett. {\bf B} 597 (2004) 39.
\bibitem{c7} BES Collaboration, J.Z. Bai et al.,
Nucl. Instrum. Methods {\bf A} 458 (2001) 627.
\bibitem{c8} S. Eidelman et al. (Particle Data Group),
Phys. Lett. {\bf B} 592 (2004) 1.
\bibitem{c9} BES Collaboration, M. Ablikim, et al.,
Nucl. Instrum. Methods {\bf A} 552 (2005) 344.
\end{thebibliography}
\end{document}